# Latitudinal Dynamics and Sectoral Structure of the Solar Magnetic Field

*Gavryuseva E.A.*
Institute for Nuclear Research RAS, prospekt 60-letiya Oktyabrya 7a, Moscow, RF, 117312
*elena.gavryuseva@gmail.com*

**Abstract.** The study of the global structure of the large-scale magnetic field of the Sun is extremely important for creating a theoretical model of the dynamics of the Sun and predictions of the real situation in the helio- and geomagnetosphere.

The purpose of the present study was to calculate the differential rotation period of a large-scale photospheric magnetic field, to study its behavior over time and to find out whether there is a sectoral structure of this field along the longitude. However, the choice of the coordinate system in which to search for it is far from unambiguous. This is closely related to the fact that the rotation of the Sun is differential in latitude and varies with depth and over time. Based on the observational data of the J. Wilcox Solar Observatory for three complete cycles of solar activity 21, 22 and 23, the period of rotation of the magnetic field at various latitudes and its change in time were calculated. A uniquely stable over 30 years longitude structure was found. It was determined that its speed of rotation coincides with the one with which the base of the convective shell rotates, that is, the structuring of the magnetic field of the Sun occurs in tachocline. This result clearly demonstrates the close connection of solar activity processes with the topology of magnetic fields, with their dynamics and depth stratification.

**Введение**

Структура и динамика Солнца определяется химическим составом, его распределением, вращением солнечной плазмы, зависящим от широты, глубины и времени (Christensen-Dalsgaard, 2002), а также от топологии крупномасштабных магнитных полей. Для адекватного понимания внутреннего строения Солнца, условий генерации ядерной энергии, потоков нейтрино и их прохождения через толщу звезды, формирования и распространения солнечного ветра и геомагнитной обстановки необходимо исследовать широкий спектр проблем, начиная с изучения его глубоких недр и кончая солнечно-земными связями. Комплексный подход к указанной проблематике последовательно реализуется. Его актуальность возрастает в последнее время в связи с ростом климатических и природных аномалий, резким увеличением запуска космических аппаратов на околоземные орбиты, исследования межпланетного пространства нашей солнечной системы. Изучение топологии и динамики крупномасштабного магнитного поля Солнца является ключевым шагом к решению этой задачи.

Для исследования этой проблемы были необходимы достаточно протяженные ряды наблюдений поверхности Солнца с хорошим разрешением по долготе и широте. Для изучения топологии и динамики магнитного поля Солнца (МПС) были использованы наблюдения фотосферного МПС на Солнечной обсерватории им. Дж. Вилкокса (The Wilcox Solar Observatory – WSO) в Калифорнии США, полученные за три цикла 21, 22 и 23 солнечной активности (СА) (Hoeksema, Scherrer, 1986).

На протяжении многих лет наиболее интенсивно изучалась солнечная активность. Было четко установлено распределение солнечных пятен (бабочки Маундера) и поведение биполярных областей согласно закону Хэйла (Hale, 1908), по которому ведущие (по солнечному вращению) и следующие за ними пятна противоположной полярности сохраняют свой знак в течение каждого 11-летнего цикла активности и противоположны в северном и южном полушариях. Полярности меняют знак от одного цикла к другому, в результате чего солнечный магнитный цикл составляет около 22 лет: во время нечетных (четных) 11-летних циклов ведущая полярность является положительной (отрицательной) в северном полушарии. Полярности высокоширотных зон противоположны и изменяются около максимумов активности (Hale, 1908). Теоретические модели, описывающие указанные закономерности, были предложены в работах (Babcock, 1955, 1961; Leighton 1969; Parker, 1955а, б). Исторический обзор эволюции МПС и механизма меридионального переноса магнитного потока с учетом дифференциального вращения был сделан в работах (Sheeley 2005; Stix, 1981). Теория солнечного динамо в дальнейшем получила бурное развитие (Charbonneau, 2020; Dikpati, Gilman, 2006; Kitchatinov, 2007; Ossendrijver, 2003; Hathaway, 2015; Соколов, 2022). Эруптивные процессы и вспышечные механизмы описаны в книге (Somov, 2020).

В классической модели цикла солнечной активности униполярные магнитные области преимущественно хвостовых полярностей переносятся к полюсам за счет меридиональных потоков и диффузии. Они постепенно гасят полярное магнитное поле предыдущего цикла. В работе (Mordvinov, Kitchatinov, 2019) авторы отметили, что смена знака МПС на полюсах имеет элемент случайности, например в 24 цикле, кроме того присутствует определенная асимметрия в активности и асинхрония между северным и южным полушариями, что существенно затрудняет прогнозирование переполюсовки. При этом не учитываются короткопериодические вариации, которые будут кратко рассмотрены в нашей статье.

Скорость дифференциального вращения собственно МПС и его вариации во времени важны для теории солнечного цикла как входящие наблюдательные данные, а также представляют самостоятельный интерес. В настоящее время накоплены разнообразные сведения о дифференциальном вращении солнечной плазмы, полученные как спектральными методами, так и отслеживанием трассеров (Beck, 2000; Paterno, 2010; Schröter, 1985; Kutsenko, Abramenko, 2022). Период вращения фотосферы на экваторе составляет около 25 дней и растет при приближении к полюсам. Методами гелиосейсмологии стало возможно определять строение и вращение в глубине Солнца

(Howe et al., 2009; Schou et al., 1994, 1998; Gavryuseva et al., 2000) и др. Благодаря решению обратной задачи по восстановлению внутреннего строения Солнца на основе расщепления частот акустических мод различными методами была установлена наиболее вероятная глубина расположения тахоклинной зоны, равная 0.715 $R_\odot$ ($R_\odot$ — радиус Солнца), которая является тонким переходным слоем от конвективной оболочки к более глубоко расположенной радиационной зоне. В глубине под тахоклином вращение становится не зависящим от широты и равным примерно 430 нГц (Schou et al., 1998). Было установлено существование так называемых торсионных колебаний, т. е. наличие бегущих к экватору волн ускорения и замедления вращения на уровне не более 1 % на низких широтах. Впервые крутильные колебания наблюдали в допплеровских измерениях потоков на поверхности (Howard, Labonte, 1980), затем было обнаружено, что они проникают в конвективную оболочку (Howe et al., 2000; Vorontsov et al., 2002; Imada et al., 2020; Javaraiah, Ulrich, 2006).

Меридиональный поток играет важную роль в перераспределении углового момента и транспортировке магнитного потока внутри Солнца. Достигнуты успехи в измерении скорости меридионального течения вблизи поверхности Солнца (около 10–15 м/с), которая варьируется с широтой и в течение цикла активности, дополнительно необходимо исследовать профиль потока в глубине (Zhao, Kosovichev, 2013). В последние годы были обнаружены волны Россби (Löptien, 2018). Накопленные сведения позволяют осуществить комплексный подход к моделированию механизма динамики Солнца, соотнося их с детальным анализом структуры крупномасштабного магнитного поля.

Для нахождения широтной структуры магнитного поля следует вычислять среднее значение МПС на каждой широте за определенный интервал времени, например за один оборот вокруг своей оси или за один оборот вокруг Солнца, т. е. за год. Такие исследования были проведены для изучения 22-летних циклов солнечной активности (Bumba, Howard, 1965, 1969; Bumba, 1976; Gavryuseva, 2005, 2006, 2006a, b, c, 2007a; Hoeksema, 1984; Obridko, Shelting, 1999; Sheeley, 2005; Solanki, 2006). В работах (Benevolenskaya, 1996, 1998; Gavryuseva, 2006, 2006a, b, c, 2007a) отмечается присутствие короткопериодических колебаний МПС, распространяющихся от экватора к полюсам.

При изучении секторной структуры обычно внимание сосредоточено на отслеживании так называемых активных долгот (Ivanov, 2007; Сулейманова, Абраменко, 2022) на видимой поверхности Солнца. Однако изучение долготной структуры МПС следует начинать с предположения о скорости ее вращения. Принято характеризовать вращение активных долгот Солнца кэррингтоновским периодом $T_{CR}$ = 27.2753 сут (Carrington Rotation – CR). Для восстановления секторной структуры МПС обычно используются свертки на кэррингтоновский интервал времени $T_{CR}$, хотя хорошо известно, что Солнце вращается дифференциально по широте и скорость вращения варьируется во времени. Поэтому для более детального изучения топологии было необходимо сначала определить периоды вращения магнитного поля фотосферы на протяжении всего интервала измерений, что и было сделано методами автокорреляции и быстрого преобразования Фурье (БПФ, англ. Fast Fourier Transform, FFT). Затем была изучена широтная и долготная структуры (ДС) в разных системах координат, включая кэррингтоновскую; дифференциально зависящую от широты, от времени, так же как МПС; и дополнительно в системе, обнаруженной из наблюдений, вращающейся с периодом $T_0$ (Gavryuseva, 2006, 2006b, e).

Расчеты долготной структуры магнитного поля были проведены на базе оригинальных данных WSO, интенсивности МПС и смоделированных полей со случайным распределением по долготе. В настоящей статье акцент сделан на вычисление периода вращения на различных широтах самого МПС и на восстановление долготной структуры МПС в системах координат, вращающихся с кэррингтоновским периодом $T_{CR}$, и с периодом $T_0$, совпадающим с вращением на том уровне, где, как предполагается, генерируется устойчивая долготная структура МПС.

Статья имеет следующую структуру. В основной части в первом разделе описаны наблюдательные данные по измерению магнитного поля Солнца.

Во втором разделе обсуждаются широтные структуры МПС с периодом 22 года и квазидвухлетние бегущие волны полярности.

Третий раздел посвящен результатам вычисления периода вращения МПС на различных широтах и его вариаций во времени.

В четвертом разделе описаны результаты изучения возможной долготной структуры МПС, вращающейся с кэррингтоновским периодом.

В пятом представлена уникально устойчивая и четко организованная на протяжении трех циклов СА долготная структура магнитного поля. В Заключении приведены наиболее значимые результаты с описанием характеристик обнаруженной устойчивой долготной структуры МПС. Обсуждается, на какой глубине она генерируется и как себя ведет в 21–23 циклах.

**Основная часть**

Вопрос о существовании глобальной структуры МПС исключительно важен как для практического предсказания геомагнитной обстановки на Земле и в ближнем космосе, так и для понимания динамики Солнца и возмущений на его поверхности. В настоящем исследовании была изучена широтная и долготная структуры, а также дифференциальное вращение фотосферного магнитного поля в течение трех циклов солнечной активности на базе данных WSO (Hoeksema, Scherrer, 1986).

**1  Наблюдательные данные**

Измерения фотосферного МПС проводились в WSO магнетометром Бэбкока по зеемановскому расщеплению спектральной линии железа FeI 525.0 нм по лучу зрения. Магнитограммы, полученные за оборот Солнца, представляют полную картину распределения магнитного поля в фотосфере за один кэррингтоновский оборот с 72 величинами МПС, равномерно распределенными по долготе $\phi$ с шагом 5° и по 30 уровням $\sin\theta$, где широта $\theta$ изменяется от 75.2° N до 75.2° S. Данные измерений располагаются на широтах $\theta$ = ±75.2, ±64.2, ±56.4, ±50.1, ±44.4, ±39.3, ±34.5, ±30.0, ±25.7, ±21.5, ±17.5, ±13.5, ±9.6, ±5.7, ±1.9 градусов. Пространственное разрешение уменьшается с увеличением $\theta$. Полярные зоны выше 75° неразрешимы. Уровень шума каждого измерения составляет менее 10 мкТл (Hoeksema et al., 1986). Измерения начаты 27 мая 1976 г., что соответствует CR1642. Анализируемый набор данных охватывает три цикла солнечной активности 21–23, начиная с минимума активности в 1976 г.

**2  Широтная структура магнитного поля**

Изучению поведения магнитного поля на разных широтах в зависимости от фазы и цикла солнечной активности было посвящено много работ (Sheeley, 2005). Накоплено большое количество информации, которая четко отражает его связь с солнечной активностью (Parker, 1955a; Solanki, 2006; Hathaway, 2015). Типичной иллюстрацией широтной структуры как функции времени является рис. 1, *a* (вверху), где четко видна четырехзонная структура с границами $\theta = 0°$ и $\theta = \pm 25°$ в максимуме солнечной активности. На этом и всех последующих рисунках положительная полярность показана красным, оранжевым и желтым цветом, а отрицательная полярность — синим и голубым. Полярности в каждой зоне чередуются и имеют противоположный знак в северном и южном полушариях. Полярность сохраняется в течение 10–11 лет, причем в приэкваториальных областях она совпадает с полярностью лидирующего пятна. Полярность высокоширотных зон запаздывает на четверть периода. Для сглаживания годовых вариаций МПС, связанных с орбитальным движением Земли вокруг Солнца, было сделано широтное усреднение магнитного поля за 1 год. Результат представлен на верхнем рис. 1, *a*. Полный период составляет ~20 лет. Эти характеристики хорошо совпадают с особенностями пятнообразовательного цикла, который показан на нижнем рис. 1, *c*. На этой кривой имеется двугорбый максимум, который отражает двухлетнюю вариацию.  Исключением является устойчивое положение границы между зонами на широтах θ = ±25° в отличие от дрейфа большинства пятен к экватору. Было проанализировано, поля какой интенсивности обусловливают формирование четырехзонной структуры. В работе (Gavryuseva, 2007а) был сделан вывод, что за нее ответственны магнитные поля средней интенсивности (5–2000 мкТл).

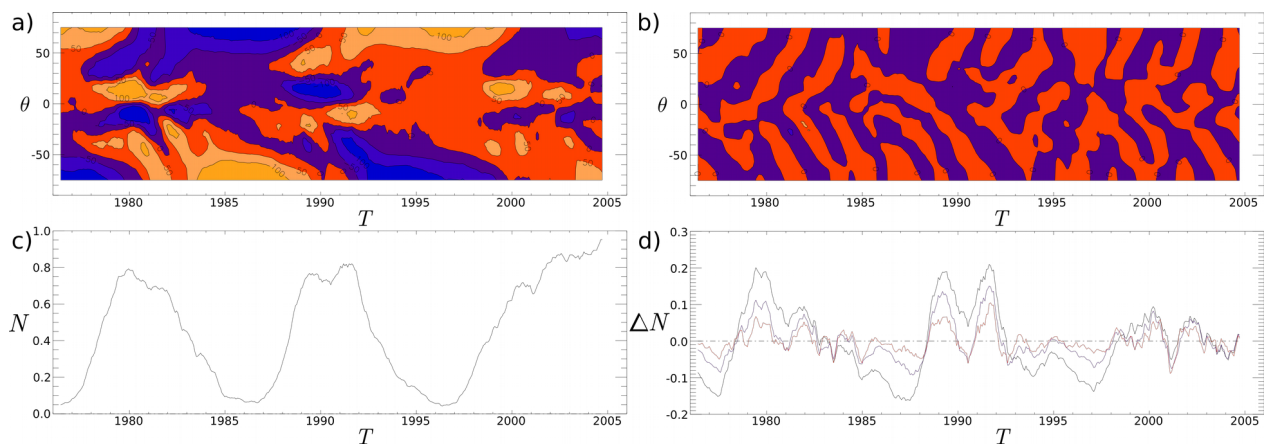

*Рис. 1, а* Магнитное поле Солнца и число солнечных пятен как функция времени в 21-23 циклах солнечной активности
    *а* — четырехзонная широтная структура при среднегодовом усреднении МПС.
    b — квазидвухлетние бегущие волны полярности магнитного поля.
    Красным и оранжевым цветом показано магнитное поле положительной полярности, синим и голубым — отрицательной. По оси Y указана широта θ в градусах, по X — время в годах.
    c — бегущее среднегодовое нормированное число солнечных пятен N.
    d — отклонения среднегодового числа пятен от среднего за 5 лет (черным), 3 года (зеленым) и 2 года (голубым цветом).

Аналогичные двухлетние вариации были отмечены в МПС (Benevolenskaya, 1996; Gavryuseva, 2006, 2006 a, c), в частоте акустических мод, измерениях радиуса и потока нейтрино (Delache et al., 1993), а также во вращении МПС (Mordvinov, Plyusnina, 2000; Hathaway, 2015).

Было проведено исследование наличия короткопериодических процессов в МПС на различных широтах в течение 21–23 циклов солнечной активности. Для этого был применен фильтр долгопериодических вариаций. На каждой широте была вычислена разница между среднегодовым и усредненным за 5, 4, 3, 2 года магнитным полем, назовем это фильтрующим вычетом (ФВ). Применяя фильтр долгопериодических вариаций было выявлено присутствие квазидвухлетних бегущих волн полярности магнитного поля на протяжении всех циклов. Наличие этих бегущих волн представлено на верхнем рис. 1, *b*. Их главное отличие от обсуждавшейся выше глобальной широтной структуры состоит в том, что они не охватывают большой интервал широт одновременно, а являются волнами, бегущими от экватора к полюсам. Это кольцо единой полярности для всех долгот фонового МПС на определенной широте, которое постепенно перемещается по широтам. Такая бегущая волна МПС доминирующей полярности распространяется от экватора к полюсам со скоростью примерно 40 км/ч (~11 м/с) и достигает полюса за ~2.5 года (Gavryuseva, 2006, 2006а, c, 2007а).

Характерным свойством бегущих волн полярности МПС является симметрия относительно экваториальной плоскости, такие волны, синхронно исходящие от экватора по направлению к полюсам, выявляются простым суммированием значений МПС на одинаковых широтах в северном и южном полушариях, которое убирает тождественные компоненты поля противоположной полярности (Gavryuseva, 2006c). Однако следует отметить, что присутствует некоторая разница в их амплитуде и периодичности, и это различие вновь приводит к синхронизации через 11 лет и 22 года (Gavryuseva, 2006а, c).

Внизу на рис. 1, *d* показано изменение со временем среднегодового нормированного числа солнечных пятен для вычетов. Квазидвухлетние вариации числа солнечных пятен присутствуют на всем протяжении кривой. Их корреляции с МПС были подробно описаны в работах (Zharkov et al., 2007a,b, 2008). Наличие такого фонового магнитного поля явно выраженной полярности (по лучу зрения) при совпадении с полярностью ведущего пятна может усиливать активность этого пятна и одновременно ослаблять величину магнитного поля и вследствие этого активность хвостового пятна (Gavryuseva, 2006; Zharkov et al., 2007a, b, 2008). Это может являться причиной квазидвухлетней периодичности пятнообразовательной активности, хорошо известной из наблюдений СА.

Переполюсовка в приполярных областях происходит примерно в максимуме цикла солнечной активности, но случается не всегда синхронно в северном и южном полушариях. Это имело место, например, в 24 цикле. Авторы работы (Mordvinov, Kitchatinov, 2019) объясняют это тем, что остатки распавшихся активных областей, дрейфующие к высоким широтам, неоднократно приносили избыточную полярность хвостового пятна к полюсам. Отсюда происходит не монотонная, а множественная смена полярности колебательного характера. Наличие квазидвухлетней периодичности бегущих волн МПС четко иллюстрирует верхний рис 1, *b*. Эти волны полярности МПС распространяются от экватора, а не возникают на средних широтах в результате распада активных областей. Их прохождение будет способствовать переносу однополярных им остатков бывших хвостовых (ведомых) областей и обусловливать квазидвухлетнюю периодичность магнитного поля вплоть до его высокоширотных вариаций.

В работе (Мордвинов, Плюснина, 2008) высказано предположение, что дрейф фонового МПС скорее представляет собой гидродинамическое явление, скорость меридионального дрейфа возмущений определяется групповой скоростью длинных баротропных волн Россби, возникающих в районе тахоклина. Авторы работ (Gilman, 2018; Kitchatinov, 2007) пришли к заключению, что при отрицательном градиенте по широте скорости дифференциального вращения возможна накачка магнитного поля под конвективной оболочкой. При анализе путей распространения звуковых волн внутри Солнца, зарегистрированных за два года с SDO/HMI были обнаружены подфотосферные меридиональные потоки. Они располагаются в три слоя по глубине: верхний движется к полюсам со скоростью 15 м/с на глубине до $0.91\,R_\odot$, второй — в более глубоком слое конвективной оболочки $(0.82–0.91)R_\odot$ со скоростью 10 м/с направлен к экватору. Еще глубже до $0.75\,R_\odot$ находится третий поток к полюсу (Zhao et al., 2013). Не исключено существование дополнительных ячеек вращения в глубине и по широте, хотя разрешение выше 60° по широте пока невозможно. Скорость распространения бегущих волн полярности магнитного поля к полюсам, представленным на рис. 2, *b*, совпадает со скоростью меридионального потока в подфотосферном слое.

### 3  Вращение фотосферного магнитного поля

Вращение Солнца имеет дифференциальный характер, изменяясь с глубиной и по широте. Большая группа ученых (Schou et al., 1998; Howe et al., 2009) методами гелиосейсмологии, решая обратную задачу по расщеплению частот акустических мод, восстановила частоту вращения и показала, что под конвективной оболочкой на дне так называемой зоны тахоклина вращение становится примерно одинаковым на всех широтах и равным 430 нГц (Schou et al., 1998; Howe, 2009).

Дифференциальное вращение плазмы фотосферы было определено различными методами, и его замедление на высоких широтах четко зафиксировано (Beck, 2000; Paterno, 2010; Kutsenko, Abramenko, 2022; Литвишко и др., 2022).

Имея данные WSO, можно вычислить скорость вращения магнитного поля. На рис. *2, a* показано изменение во времени периода дифференциального вращения магнитного поля в южном и в северном полушариях на широтах −75°÷+75° в 21–23 циклах солнечной активности. Явно виден дифференциальный по широте характер вращения крупномасштабного МПС. Синим и голубым цветом показано более быстрое вращение приэкваториальных зон, а красным и желтым — более медленное на высоких широтах (период растёт от 24.5 до 31 суток). Период вращения *P* варьируется во времени, зависит от интенсивности магнитного поля, уменьшается при ослаблении поля. В приполярных областях интенсивность магнитного поля возрастает в 1976, 1985 и 1995 годах во время минимумов солнечной активности. При этом период вращения постепенно увеличивается с широтой, достигая замедления вращения вплоть до 1.5 дней на широтах от +/-60° до

+/-75°. Во время переполюсовки, которая происходит в приполярных зонах в максимумы активности примерно в 1980, 1990 и 2000 годах, вращение ускоряется.

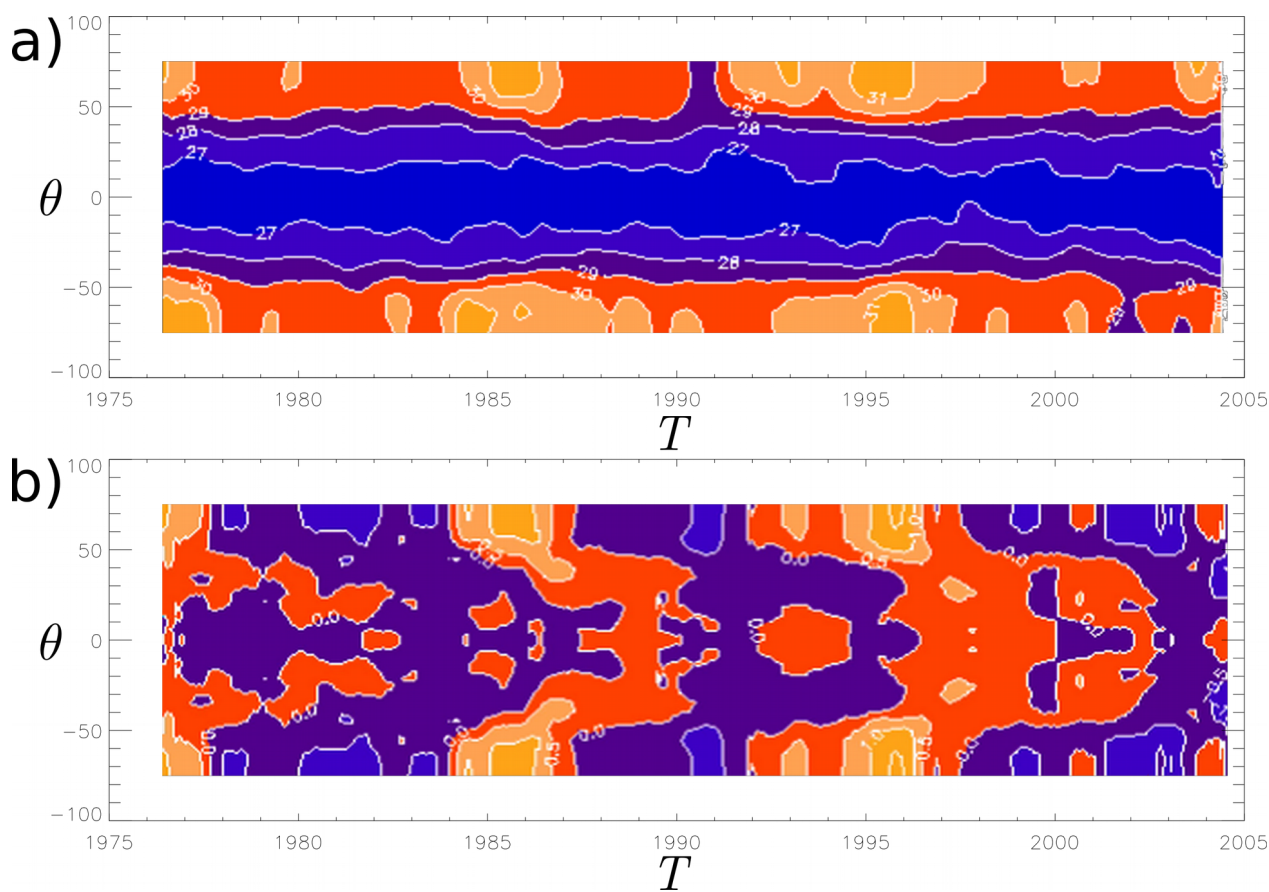

*Рис. 2* Период вращения магнитного поля Солнца и его отклонение от среднего как функция времени и широты

*a* — период дифференциального вращения магнитного поля в южном и в северном полушариях как функция времени в годах в 21–23 циклах солнечной активности. Контуры соответствуют периодам 27, 28, 29, 30 и 31 сут (голубого, синего, красного, оранжевого, желтого цвета).

*b* — среднее отклонение (север–юг) периода вращения, полученное FFT-анализом подмножеств длительностью 40CR, от среднего периода за 1976–2004 гг. на каждой широте. Синий, голубой (красный, оранжевый, желтый) цвета соответствуют отрицательному (положительному) отклонениям, контуры — отклонениям периода вращения равным 0, ±0.5, ±1.0, ±1.5 сут.

На рис. 2, *b* представлено среднее по север–юг отклонение периода вращения *P*, полученное FFT-анализом подмножеств длительностью 40CR (около 3 лет) от среднего периода за 1976–2004 гг. Этот интервал соответствует циклам 21 и 22 (выбран полный 22-летний период во избежание возможного влияния изменчивости циклов активности). Средний период дифференциального вращения также вычислялся как среднее по периодам вращения, рассчитанным на каждой широте для более коротких подмножеств длительностью 40CR.

На рис. 2, *b* видны бегущие волны ускорения и замедления вращения МПС, распространяющиеся от полюсов к экватору. Периодичность этих волн ~11 лет. Ускорение вращения совпадает со сменой полярности, когда магнитное поле на полюсах слабое, что согласуется с результатами (Mordvinov, Plyusnina, 2005; Obridko, 2001). Недавние результаты анализа десяти лет наблюдений SDO/HMI с высоким разрешением трассеров в 24-м цикле подтверждают, что при более слабом МПС скорость вращения увеличивается, и меридиональные потоки ускоряются (Imada, 2020).

Если на широтах выше 20 градусов присутствует только квазиодиннадцатилетняя периодичность, то в приэкваториальной зоне имеется указание и на крутильные колебания с периодом ~5 лет (Gavryuseva, 2006, 2006d). Торсионные волны, впервые обнаруженные в 1980 г. (Howard, Labonte, 1980) на средних и низких широтах, сейчас находятся под пристальным изучением благодаря наблюдениям из космоса с высоким разрешениям (SOHO, SDO).

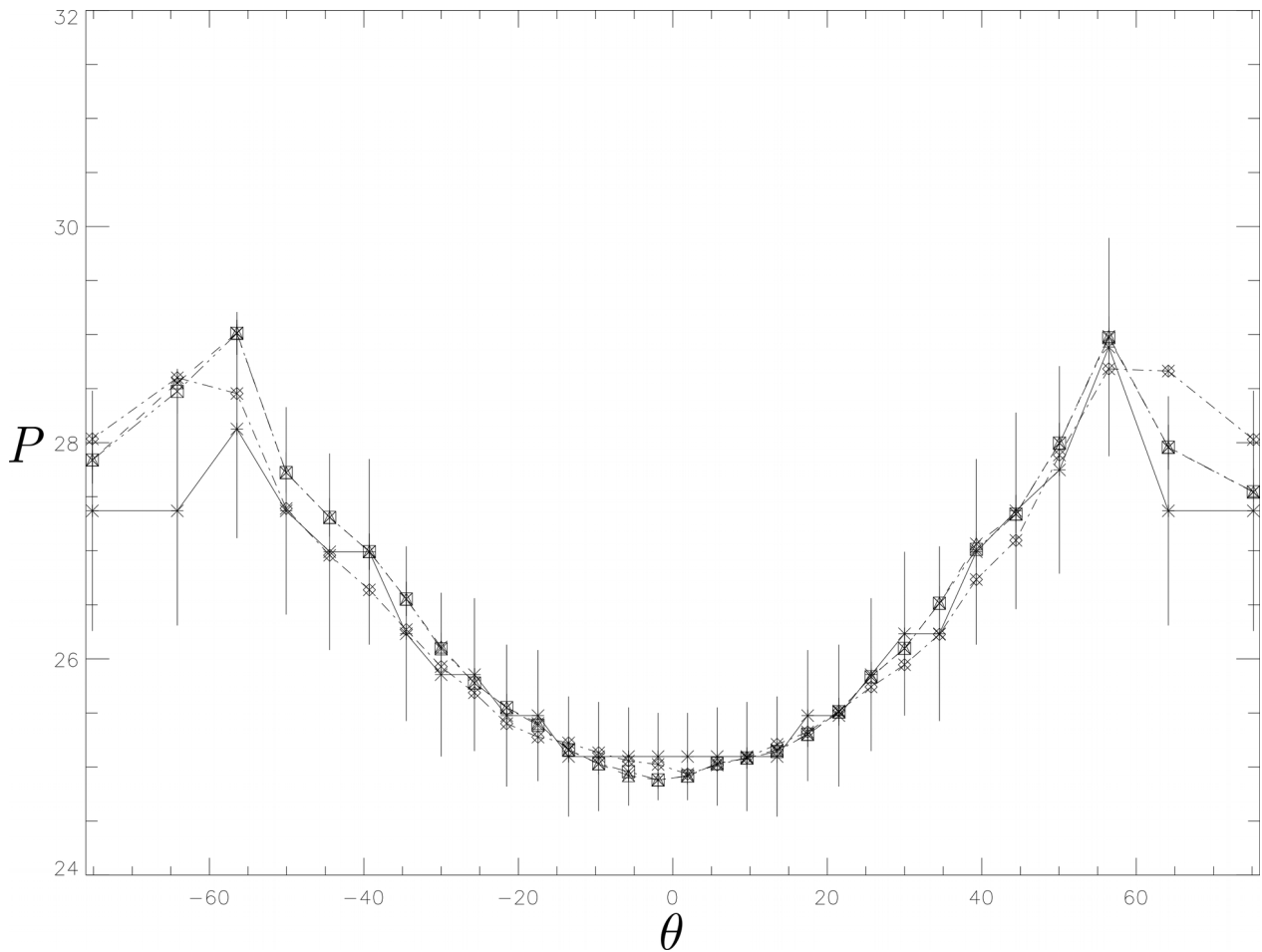

*Рис.3* Период дифференциального вращения как функция широты θ, в градусах

Непрерывная линия - сидерический период вращения МПС (в сутках) с ошибками, полученный авто-корреляционным методом. Пунктирная и точечная линии - периоды вращения плазмы, рассчитанные FFT методами.

Средний за 22 года сидерический период дифференциального вращения МПС как функция широты показан на рис. 3. Непрерывной линией отмечен сидерический период вращения МПС, полученный автокорреляционным методом полного ряда данных WSO на каждой широте. Штриховая и точечная линии — периоды вращения плазмы, рассчитанные методами FFT для полного ряда данных и для среднего по периодам, вычисленным для его подмножеств, соответствующих отдельным циклам.

Сидерический период вращения $P$ крупномасштабного магнитного поля совпадает с результатами спектроскопических измерений вращения фотосферной плазмы в интервале широт $θ = -45°÷45°$. На рис. 3 показано, что на широтах 55°÷60° период вращения $P$ достигает максимальной величины 29 дней, а на более высоких широтах вращение МПС квазитвердотельное. При этом различие между периодами вращения МПС и фотосферной плазмой достигает 2÷3 дней. Такой результат (Gavryuseva, 2006, 2006d) находится в согласии с широтной зависимостью периода вращения, полученной на основе данных Национальной обсерватории Китт-Пик по наблюдениям 1959–1985 гг. (Stenflo, 1989). Этот интересный результат можно объяснить всплытием на поверхность на широтах 55°–60° магнитного поля со дна конвективной оболочки, где в тахоклине вращение не зависит от широты и близко к тому, что регистрируется по фотосферному МПС на $θ = 55°–60°$ (Howe, 2009).

В отличие от определения периода вращения солнечной плазмы спектроскопическими методами было обнаружено, что на широтах выше $θ = ±60°$ в северном и южном полушариях вращение магнитного поля более не замедляется, а остается на уровне, соответствующем скорости вращения Солнца под основанием конвективной оболочки ~430 нГц, которая была вычислена несколькими группами ученых по расщеплению частот акустических колебаний методами гелиосейсмологии (Howe et al., 2009). Этот факт можно рассматривать как указание на то, что формирование глобального магнитного поля связано с глубинными слоями под конвективной оболочкой. Это полезно учитывать при разработке динамо-моделей. Дополнительным аргументом в пользу этого является вывод, к которому пришли авторы, анализируя измерения прибора MDI, что изменение в меридиональном потоке, вероятно вызвано изменением магнитного поля от минимального до максимального значения вблизи тахоклина, что, в свою очередь, указывает на то, что там и происходит генерация МПС (Zhao et al., 2013).

**4 Долготная структура магнитного поля, вращающаяся с кэррингтоновским периодом**

Секторная структура гелиосферы, условия распространения солнечного ветра и геомагнитные возмущения тесно связаны с долготной структурой (ДС) магнитного поля, а также с вращением и активностью Солнца. В связи с этим изучение ДС МПС важно с практической и теоретической точек зрения. Однако вопрос, существует ли достаточно устойчивая и четкая долготная структура магнитного поля, остается открытым. Если существует, необходимо понять, с какой скоростью она вращается, насколько долгоживущей является, на какой глубине формируется.

Если долготное структурирование МПС происходит в глубоких недрах и жестко привязано к вращению на уровне генерации, следует ожидать проявление долготной структуры, вращающейся таким же образом, как тот слой, где она формируется (Gavryuseva, 2006e, 2007b). Если дифференциальное вращение верхних слоев оказывает решающее влияние на фотосферное распределение МПС, то широтная зависимость вращения должна проявляться и в долготной структуре, чья скорость вращения будет иметь величину из диапазона фотосферных значений. В этом случае реконструкцию ДС следовало бы выполнять с учетом характера вращения фотосферного МПС. Поскольку мы знаем, что эта скорость вращения изменяется во времени, система координат должна также следовать этому изменению. Результаты такой реконструкции можно найти в работе (Gavryuseva, Godoli, 2006), где были восстановлены долготные структуры МПС на разных широтах в системе координат, вращающейся дифференциально и с переменной скоростью во времени, именно такой, как она представлена на рис. 2, *а*. Для шести интервалов широт в такой системе координат, которая вращается как магнитное поле Солнца, дифференциально по широте и с вариациями во времени, были восстановлены долготные структуры. На приэкваториальных широтах до θ = ±22° долготная структура демонстрирует явно выраженный наклон под углом ~85° при ее смещении во времени. Этот факт указывает на то, что вращение самой ДС более медленное, чем реальное вращение МПС фотосферы и чем вращение активных широт с характерным периодом $T_{CR}$. Поскольку долготная структура не вращается так, как магнитное поле фотосферы, то следовательно, ее формирование может происходить в более глубоких слоях Солнца.

Обычным подходом к изучению этой проблемы является восстановление долготного распределения путем свертки на период кэррингтоновского вращения, характерный для активных широт. Такой подход был успешно применен многими авторами, использующими как МПС, так и его интенсивность, солнечные пятна и другие трассеры (Hathaway, 2015; Obridko et al., 2011). Результаты, полученные таким путем, схожи в том, что выявляется наличие двух сегментов, отстоящих на ~150°, где отмечен более высокий уровень МПС или активности. Мы провели аналогичные исследования, рассчитали долготные структуры как свертки на интервал времени, равный $T_{CR}$, рядов наблюдений МПС на всех широтах в 21-23 циклах как для интенсивности, так и для магнитного поля с учетом его полярности (Gavryuseva, Godoli, 2006; Gavryuseva, 2006e, 2007b).

На рис. 4, *а* слева представлена долготная структура измеренного магнитного поля в координатной системе X—Y, где по осям отложены долгота ϕ и широта θ, в градусах), вращающейся с периодом Кэррингтона $T_{CR}$. Заметно присутствие двух секторов более выраженной магнитной интенсивности, разнесенных на 150°. Возникает вопрос, может ли подобная структура сформироваться при случайном долготном распределении МПС. Для того, чтобы исследовать эту возможность, было проведено численное моделирование МПС с учетом процессов солнечной активности со случайным

ДС. Это было сделано простым способом. Для каждого CR оборота проводился случайный розыгрыш долготного местоположения магнитного поля из реальных величин МПС этого CR оборота и затем проводилась свертка с периодом $T_{CR}$.

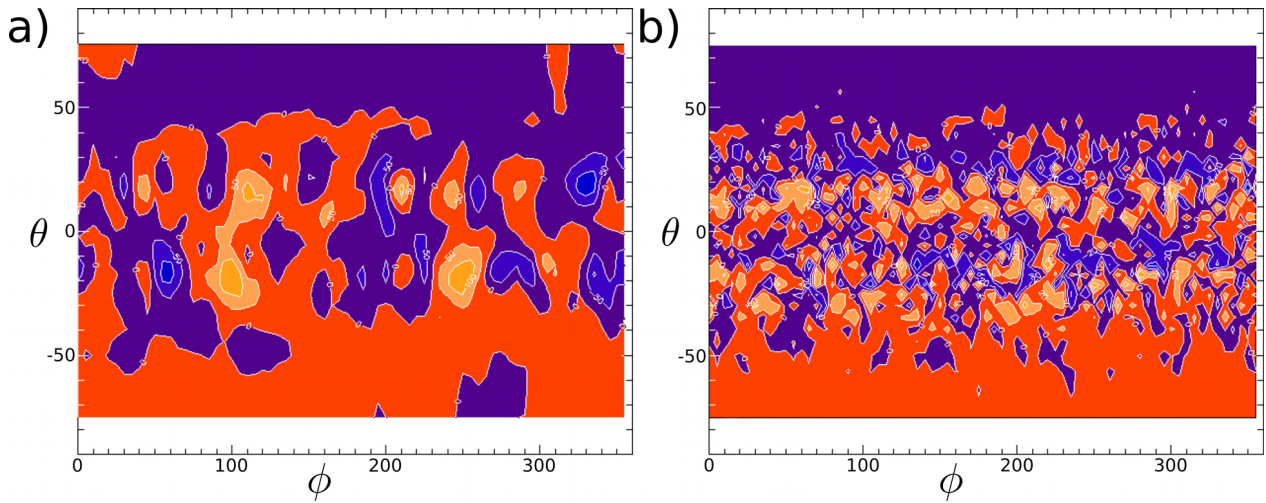

*Рис. 4* Долготная структура магнитного поля, вращающаяся с кэррингтоновским периодом $T_{CR}$.
*a* — для реального фотосферного магнитного поля; *b* — для смоделированного случайного долготного расположения МПС. Красным, оранжевым, желтым цветом показано магнитное поле положительной полярности, синим и голубым — отрицательной полярности. Долгота $\phi$ и широта $\theta$ в градусах.

Усредненное результирующее долготное распределение показано на рис. 4, *b* для модели долготного распределения случайного характера, но при характеристиках величины и временной зависимости МПС, аналогичной реальной, т. е. с учетом циклов активности. Очевидна разница, поэтому можно сделать однозначный вывод, что чисто случайное долготное распределение не соответствует ДС реального МПС (Gavryuseva, Godoli, 2006).

Еще одной важной характеристикой долготных структур является их устойчивость. Для изучения этого вопроса были вычислены усредненные ДС при свертке на $T_{CR}$ (то есть в предположении о их вращении с периодом $T_{CR}$) коротких подмножеств WSO данных на каждой широте $\theta$. Таким образом было выявлено временное поведение долготной структуры при свертке на $T_{CR}$ в 21–23 циклах СА. На рис. 5, *a* показано изменение ДС во времени, по оси X отложено время $T$ в годах, а по оси Y — долгота $\phi$ в днях (в пересчете на вращение с периодом $T_{CR}$, то есть полный оборот 360° совершается за период $T_{CR}$). Красным и оранжевым (синим и голубым) цветом показано поле положительной (отрицательной) полярности на разных долготах $\phi$ как функция времени $T$.

Следует отметить, что, как продемонстрировано на рис. 5, *a* существуют узкие долготные сектора около седьмого и шестнадцатого дня от начала полного $T_{CR}$ оборота, где полярность сохраняет свой знак в течение длительного времени, вплоть до полного цикла СА. Но особой четкой структуры по долготе и/или симметричной в обоих полушариях не наблюдается. Интегральные свертки ДС за 21–23 циклы СА показаны на рис. 5, *b*. Сплошной, штриховой и штрих-пунктирной линиями показаны свертки долготных структур МПС, усредненные по всем широтам, отдельно в северном и отдельно в южном полушариях. Некоторое указание на регулярность присутствует, но слабо выраженное. Это может иметь место при определенной соразмерности периода вращения ДС, свертки и внутренних характеристик самой долготной структуры. Такой результат будем считать указанием на то, что некая ДС существует, но следует приложить дополнительные усилия для ее обнаружения (Gavryuseva, 2006e).

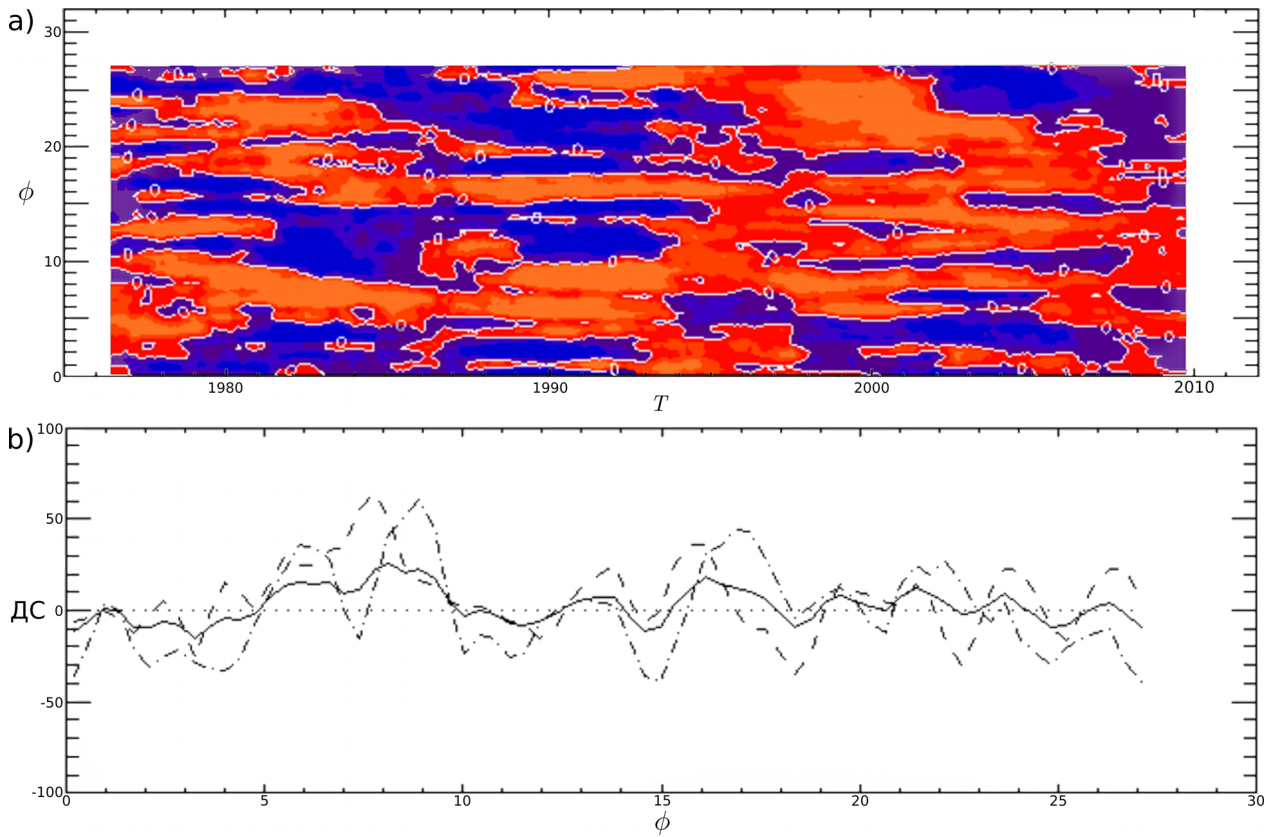

*Рис. 5* Динамика долготной структуры магнитного поля в кэррингтоновской системе
*а* — изменение во времени ДС при свертке магнитного поля на интервал $T_{CR}$=27.2753 сут (ось X — время в годах, ось Y — долгота в градусах), поле положительной и отрицательной полярности показано красным и синим цветом; *b* — долготная структура магнитного поля (в мкТл), как функции долготы (ось X в днях, в пересчете на оборот $T_{CR}$), усредненные за 21–23 циклы СА по всем широтам, по северному и по южному полушариям (сплошная, штриховая и штрих-пунктирная линии).

## 5 Устойчивая четко организованная секторная структура магнитного поля

Имея лишь указания на возможное существование ДС, мы не знаем априори, с какой скоростью она вращается. Предыдущие результаты указывают на то, что ДС может формироваться в глубине Солнца. При этом ее вращение может происходить с той же скоростью, с какой происходит ее формирование. Изначально не известно, где она генерируется, если это вообще имеет место быть. Самым лучшим было бы получить ответ на этот вопрос из реальных наблюдательных данных, а не строить предположения. Математически значимым указанием и подсказкой для дальнейшего обнаружения ДС МПС стал расчет коэффициента автокорреляции $K_а$ рядов магнитного поля по WSO измерениям на всех широтах θ.

На рис. 6 показан коэффициент автокорреляции МПС при смещении рядов данных на каждой широте на 800 градусов вдоль оси X, то есть по долготе φ на два оборота. Можно сделать два вывода. Во-первых, примерно через один оборот, затем и через второй, четко видно общее возрастание коэффициента автокорреляции, причем увеличение $K_а$ запаздывает на более высоких широтах, что соответствует дифференциальному характеру вращения по широте. Однако это замедление вращения продолжается лишь до широт 55°–60°, а затем останавливается или даже немного спадает, что наглядно подтверждает результат, представленный в разделе 3. Такая зависимость периода вращения глобального МПС на широтах выше 50° согласуется с результатами работ (Stenflo, 1989; Obridko, Shelting, 2001) о более жестком вращении приполярного магнитного поля и корональных дыр.

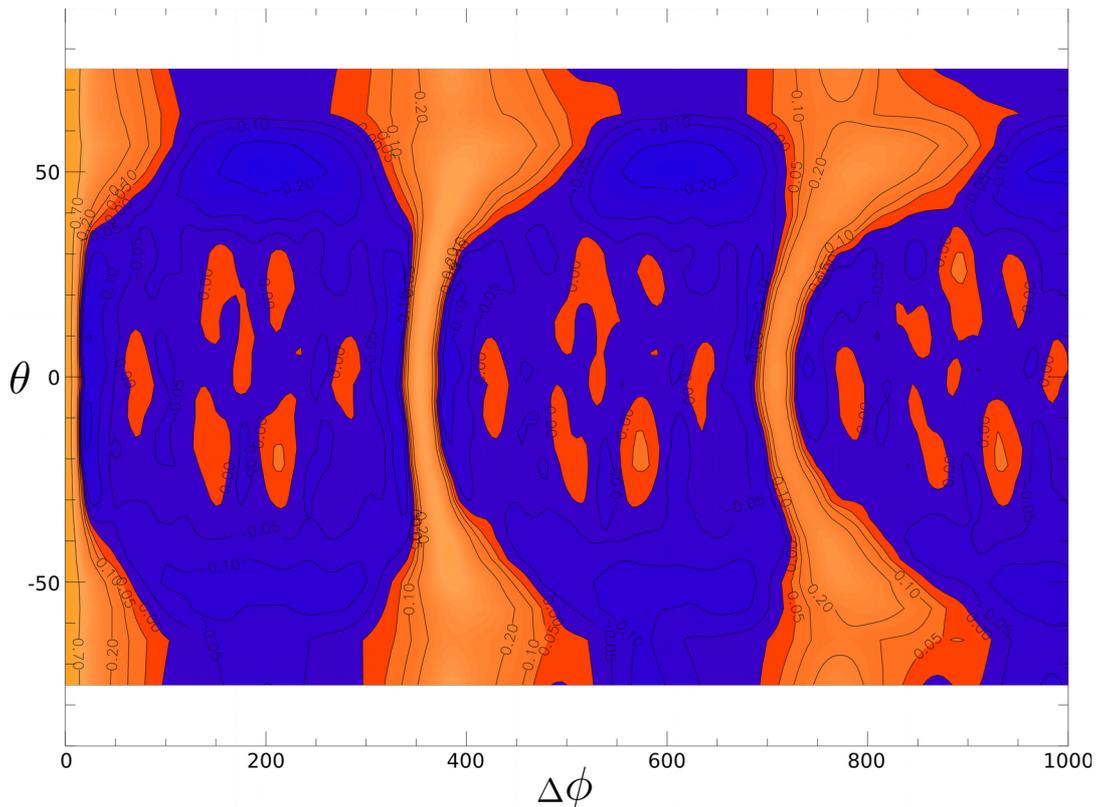

*Рис. 6* Автокорреляция магнитного поля как функция смещения по долготе (вдоль оси X в градусах) на разных широтах (ось Y в градусах)

Второй интересной особенностью является присутствие четких пиков автокорреляции на 1/5 и 4/5 долях на экваторе и на 2/5 и 3/5 долях от периода вращения на θ = ±25°, что является отражением наличия скрытой структуры. Этот результат был тщательно проанализирован и позволил установить период вращения $T_0 \approx 30.321$ сут, который предположительно мог бы соответствовать скорости вращения ДС. Были проведены расчеты по восстановлению долготной структуры при вращении с периодом $T_0$. Результаты свертки именно с этим периодом $T_0$ на разных широтах за три цикла активности показаны на рис. 7.

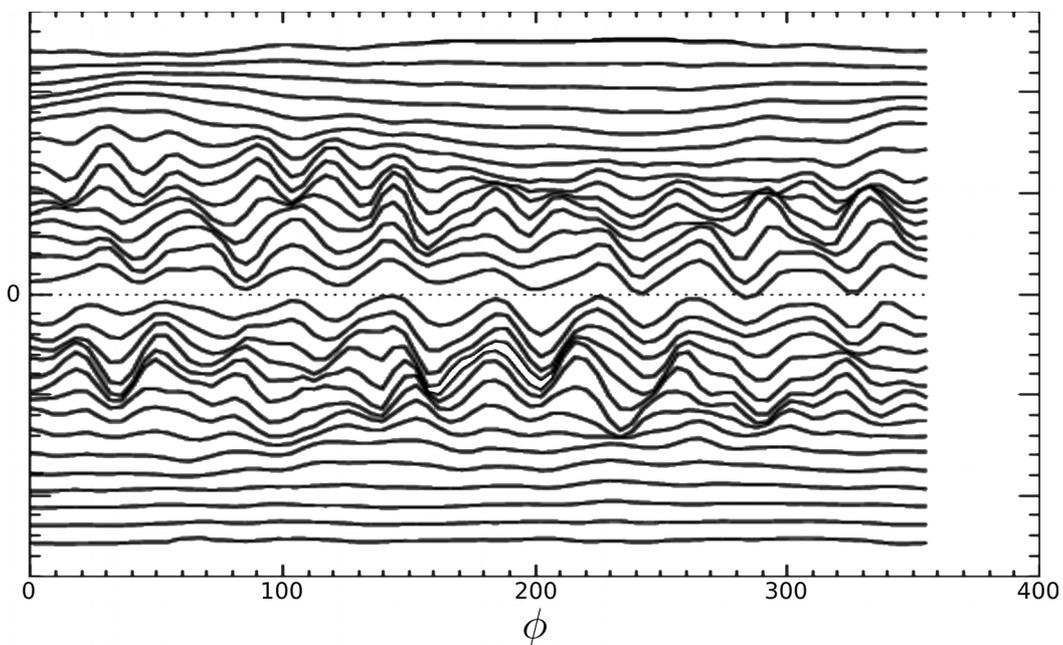

*Рис. 7* Долготная структура магнитного поля при свертке на период $T_0 = 30.321$ сут.
Для 30 разных широт (снизу вверх) от –75° до +75° (без привязки к оси Y) показана долготная зависимость, средняя за три цикла 21–23 при свертке на $T_0$ (предполагаемый период вращения зоны ее генерации).

По оси X отложена долгота φ в градусах. Одна над другой показаны долготные структуры на каждой из 30 широт, где производились измерения, без привязки их реального положения к оси Y. Это сделано для наглядной демонстрации реальной формы ДС на каждой широте в отдельности, для визуализации регулярности периодической формы ДС и показа, на каких широтах эта регулярность имеет место. Как следует из Рис. 7 долготные структуры четко представлены на всех средних широтах в обоих полушариях примерно вплоть до θ = ± 40° и симметричны в обоих полушариях.

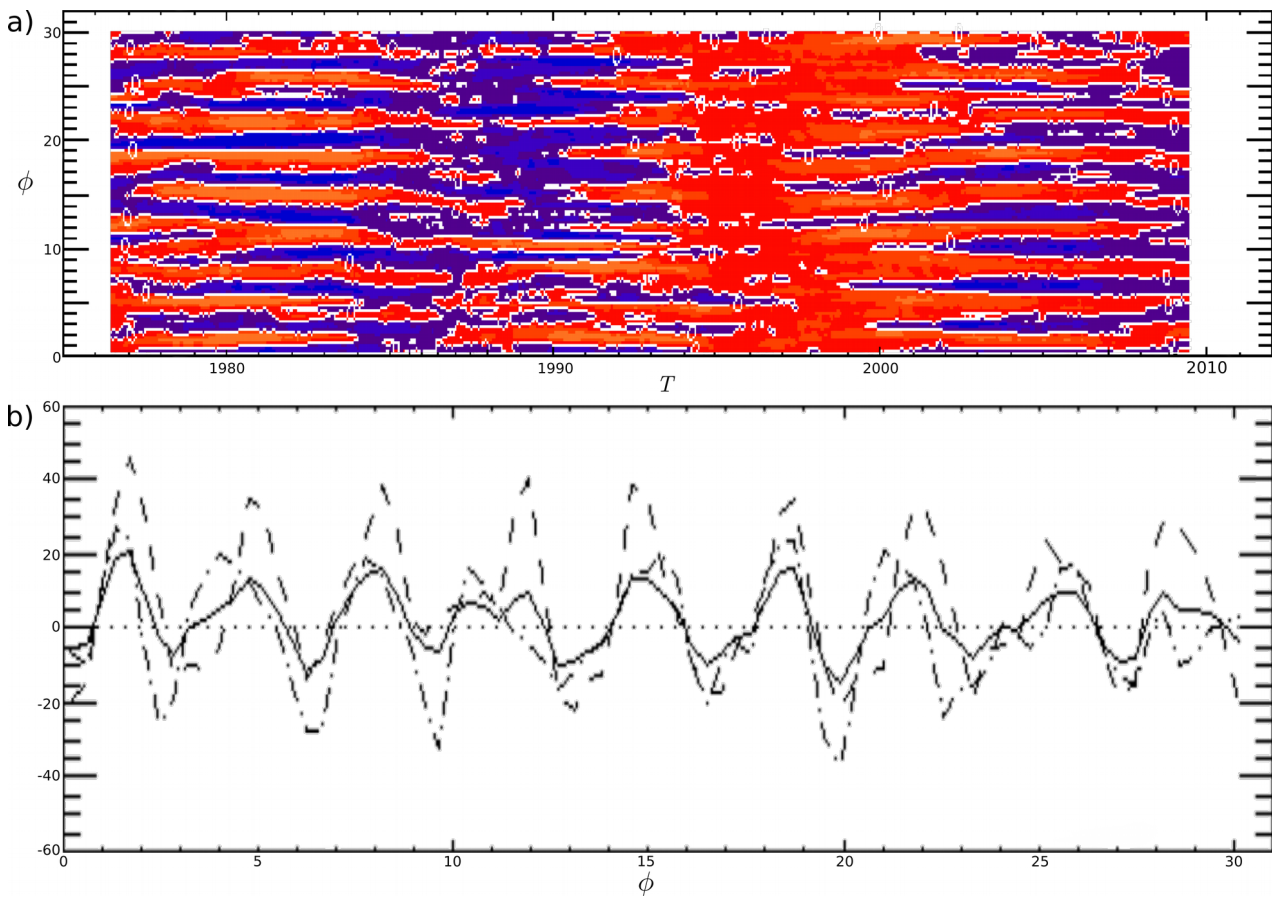

*Рис. 8* Динамика долготной структуры в системе координат, вращающейся с периодом $T$o

*a* — изменение во времени ДС при свертке магнитного поля на интервал $T$o (ось X — время в годах, ось Y — долгота в днях, в пересчете на период оборота $T$o), поле положительной и отрицательной полярности показано красным и синим цветами; *b* — долготные структуры магнитного поля (в мкТл), как функции долготы (ось X в днях, в пересчете на оборот $T$o), усредненные за 21–23 циклы СА по всем приэкваториальным широтам, по северному и по южному полушариям (сплошная, штриховая и штрих-пунктирная линии).

Важным вопросом является устойчивость обнаруженной долготной структуры во времени, от цикла к циклу и внутри цикла солнечной активности. Для более пристального отслеживания ее поведения во времени на каждой широте была проведена свертка магнитного поля на интервал времени, равный $T$o на подмножествах длительностью 1 год. Бегущим средним интервал перемещался во времени по ряду данных наблюдений в 21–23 циклах. Для того, чтобы уменьшить фоновое влияние широтной структуры и её циклических вариаций с солнечной активностью, следует предварительно отфильтровать широтную структуру. На рис. 8, *а* показано изменение долготной структуры с годами при свертке на интервал $T$o. Имеет место устойчивое сохранение полярности в каждом секторе долготы, сектора разной полярности регулярно чередуются, границы секторов устойчивы на протяжении солнечного цикла. Сравнение с аналогичным рис. 5, *а*, соответствующим временному поведению долготной структуры, вращающейся с периодом $T_{CR}$, безусловно подтверждает более высокую и устойчивую организацию ДС, вращающейся с периодом $T$o.

Суммарная свертка по всем приэкваториальным широтам, а также отдельно для северного и для южного полушарий показана на рис. 8, *б*. Все эти долготные структуры имеют периодический характер. Сектора полярности одинакового знака отстоят друг от друга на 40° в системе координат, вращающейся с периодом $T$o. Наличие определенной долготной организации в кэррингтоновской системе координат легко объясняется соразмерностью 40-градусной периодичности самой долготной структуры с периодами вращения $T_{CR}$ и $T_0$.

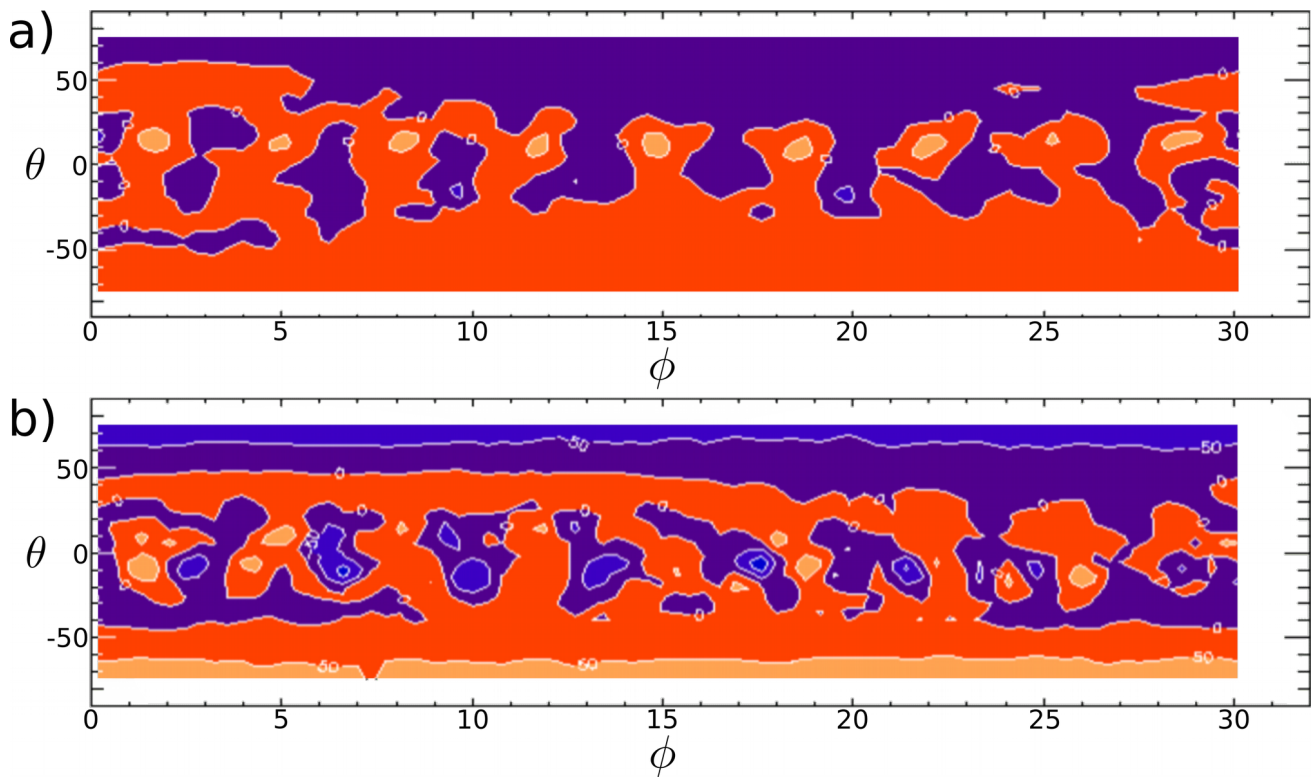

*Рис. 9* Устойчивая долготная структура, вращающаяся с периодом $T_0$.
*а* — долготно-широтная (по XY осям) развертка магнитного поля Солнца, усредненная за 21–23 циклы; *b* — развертка, усредненная только за период минимума солнечной активности 23 цикла. Долгота по оси X в днях, в пересчете на вращение с периодом $T_0$; широта по оси Y в градусах.

Таким образом приходим к заключению, что существует исключительно устойчивая на протяжении трех циклов солнечной активности долготная структура, состоящая из секторов чередующейся полярности, сохраняющая фазу от цикла к циклу и симметрично присутствующая в обоих полушариях. Эта структура тем лучше проявляется, чем длиннее ряд наблюдений. При добавлении новых данных 23-го цикла к двум предыдущим, т. е. при удлинении ряда, ДС становилась все более четкой, не изменяя при этом свою форму.

Результирующее долготно-широтное распределение магнитного поля при вращении с периодом $T_0$, усредненное за 21–23 циклы, представлено на рис. 9, *а* (вверху). Обращает на себя внимание симметрия север–юг, равномерность расположения и форма секторной структуры магнитного поля.

Эти результаты могут быть объяснены генерацией и формированием четкой ячеистой (секторной) структуры в глубине Солнца под основанием конвективной оболочки, где вращение примерно одинаково на всех широтах и происходит с периодом близким к $T_0$.

Процессы, происходящие в фотосфере, во многом случайного характера, часто мелкомасштабные, но интенсивные, могут маскировать регулярную глубинную долготную структуру МПС, особенно при высокой солнечной активности. Поэтому было интересно проверить, присутствует ли эта уникальная структура в условиях спокойного Солнца. Интервал минимума солнечной активности 23-го цикла идеально подходил для такого рода проверки. На базе наблюдательных данных за интервал минимума СА 23-го цикла была рассчитана ДС с периодом вращения $T_0$. На рис. 9, *б* представлена долготно-широтная развертка магнитного поля, рассчитанная с периодом вращения $T_0$, усредненная только за интервал минимума солнечной активности 23-го цикла. Благодаря ослаблению влияния случайных возмущений на видимую картину МПС, ДС была четко выявлена даже за столь короткий интервал времени. Это можно рассматривать как подтверждение глубинной природы долготной устойчивой секторной структуры магнитного поля Солнца.

## ЗАКЛЮЧЕНИЕ

Изучение топологии и динамики глобального магнитного поля Солнца было проведено на базе наблюдательных данных WSO, покрывающих три цикла солнечной активности 21–23. В этой работе было установлено, что широтная зависимость представляет собой комбинацию четырехзонной структуры с 20–22-летним периодом и бегущих со скоростью около 40 км/ч от экватора к полюсам волн полярности фонового магнитного поля с периодом 2–3 года. В свете последних сведений о меридиональных потоках, которые под фотосферой движутся к полюсам примерно с такой же скоростью (Zhao et al., 2013; Imada, 2020), естественно полагать, что такие процессы связаны с бегущими волнами полярности и при взаимодействии с дифференциальным вращением будут иметь периодический характер. В этой связи кажутся важными усилия, предпринимаемые теоретиками (Pipin, Kosovichev, 2020; Gilman, 2018; Kitchatinov, 2007) и исследователями потоков внутри Солнца.

Вычислен период дифференциального вращения МПС. Вращение замедляется по мере удаления от экватора до широт 55°–60°, а на более высоких широтах остается примерно постоянным и равным скорости вращения тахоклинной зоны. Вычислен период вращения магнитного поля на разных широтах и найдено его изменение во времени с периодом 11 лет в виде бегущей волны, распространяющейся от высоких широт к экватору, а также его вариация с периодом около 5 лет в приэкваториальной зоне. Это говорит о том, что магнитные поля, регистрируемые на приполярных широтах, связаны с глубинными слоями Солнца. Скорость всплывания магнитного поля должна быть достаточно быстрой, чтобы обеспечить наблюдаемый эффект. Технически разрешение процессов в приполярных зонах очень затруднено, но становится все более важным.

Долготная структура магнитного поля была вычислена в предположении вращения с кэррингтоновским периодом, подтверждено присутствие в ней двух несимметрично расположенных активных секторов, разнесенных на 150°. Сравнение с рассчитанной моделью случайного долготного распределения показало, что ДС реального магнитного поля имеет неслучайный характер. Это может быть обусловлено соразмерностью долготной структуры Солнца, имеющей период 40°, с кэррингтоновским периодом $T_{CR}$.

Автокорреляция длинных рядов наблюдений магнитного поля выявила наличие скрытой долготно-широтной структуры. Был определен период ее вращения $T_0 \approx 30.321$ сут, который показывает, что зона формирования этой структуры может находиться в тахоклине, где вращение происходит примерно с таким же периодом $T_0$, одинаковым для всех широт. Восстановлена ДС магнитного поля, вращающаяся с периодом $T_0$, и изучено ее поведение в каждом цикле солнечной активности в северном и в южном полушариях. Она устойчива на протяжении трех циклов солнечной активности, симметрична в обоих полушариях, а долготные сектора сохраняют фазу от цикла к циклу.

Присутствие долготной секторной структуры, вращающейся с периодом $T_0$, было обнаружено на интервале времени, соответствующем минимуму солнечной активности 23-го цикла. Это подтверждает то, что эта структура не является продуктом длительного присутствия отдельных активных групп пятен, а имеет глубинную природу. Синусоидальная форма этой ДС и ее симметричное присутствие в обоих полушариях Солнца говорит о неслучайном характере ее формирования в конвективной оболочке. Для определения механизма формирования нужно провести как теоретическое моделирование, так и продолжить ее изучение в последующих циклах солнечной активности.

В свете полученных результатов подтверждается, что Солнце является сложной системой, в которой происходят гармонично связанные между собой динамические процессы, имеющие периодический характер и представляющие интерес для дальнейшего изучения.